# 3D-printed axicon enables extended depth-of-focus intravascular optical coherence tomography

Pavel Ruchka[1], Alok Kushwaha[2,3], Jessica A. Marathe[4-6], Lei Xiang[2,3], Rouyan Chen[2,3,6], Rodney Kirk[3,4], Joanne T. M. Tan[4,6], Christina A. Bursill[4,6], Johan Verjans[4-6], Simon Thiele[7], Robert Fitridge[4,8], Robert A. McLaughlin[3,4], Peter J. Psaltis[4-6], Harald Giessen[1], and Jiawen Li[2,3,6]

[1] 4th Physics Institute and Research Center SCoPE, University of Stuttgart, 70569 Stuttgart, Germany
[2] School of Electrical and Mechanical Engineering, University of Adelaide, Adelaide, SA 5005, Australia
[3] Institute for Photonics and Advanced Sensing, University of Adelaide, Adelaide, SA 5005, Australia
[4] Faculty of Health and Medical Sciences, University of Adelaide, Adelaide, SA 5005, Australia
[5] Department of Cardiology, Central Adelaide Local Health Network, Adelaide, SA 5000, Australia
[6] Lifelong Health Theme, South Australian Health and Medical Research Institute (SAHMRI), Adelaide, SA 5000, Australia
[7] Institute of Applied Optics (ITO) and Research Center SCoPE, University of Stuttgart, 70569 Stuttgart, Germany
[8] Vascular and Endovascular Service, Central Adelaide Local Health Network, Adelaide, SA 5000, Australia

**Abstract**

A fundamental challenge in endoscopy is how to fabricate a small fiber-optic probe that can achieve comparable function to probes with large, complicated optics (e.g., high resolution and extended depth of focus). To achieve high resolution over an extended depth of focus (DOF), the application of needle-like beams has been proposed. However, existing methods using miniaturized needle beam designs fail to adequately correct astigmatism and other monochromatic aberrations, limiting the resolution of at least one axis.

Here, we describe a novel approach to realize freeform beam-shaping endoscopic probes via two-photon direct laser writing, also known as micro 3D-printing. We present a design achieving approximately 8 µm resolution with a DOF of >800 µm at a central wavelength of 1310 nm. The probe has a diameter of 250 µm (without the catheter sheaths) and is fabricated using a single printing step directly on the optical fiber. We demonstrate our device in intravascular imaging of living atherosclerotic pigs at multiple time points, as well as human arteries with plaques *ex vivo*.

This is the first step to enable beam-tailoring endoscopic probes which achieve diffraction-limited resolution over a large DOF.

**Introduction**

Life-threatening diseases, such as coronary artery and cerebrovascular diseases, are often indicated to the presence of atherosclerotic plaques. Optical coherence tomography (OCT) is a promising technique for *in vivo* assessment of atherosclerotic plaques. In particular, intravascular OCT can characterize high-risk atherosclerotic plaques, is commonly used in coronary arteries to help diagnose high-risk plaques, and potentially guide treatment to prevent heart attacks[1,2].

*In vivo* intravascular imaging inside small to moderate sized arteries, such as coronaries, enforces extremely limiting restrictions on fiber-optic OCT probes. The focusing optics of the OCT probe must be small enough to image safely inside a narrow blood vessel, which limits the sophistication of the optical design. The probe is encased within a protective catheter, which also introduces optical aberrations, such as astigmatism[3,4]. Whilst current commercially available probes have demonstrated clinical value, they fail to exploit the cellular-level diagnostic potential of OCT[5]. A key challenge is to enable high-resolution intravascular imaging over an extended depth of focus (DOF)[6].

Needle-beams (also called Bessel beams) are one approach designed to maintain high resolution over a greater depth than is possible with conventional Gaussian beams. Needle-beams are widely used in advanced microscopy, including OCT[7-9], light sheet microscopy[10,11], multiphoton microscopy[12,13], and photoacoustic microscopy[14,15], due to their extended depth of focus with small lateral extension of the beam. Various approaches have been proposed to achieve needle-shaped beams for extended DOF in OCT endoscopes, including: axicon lenses fabricated by molding a fiber[16], spatial filters[17,18] or annual apodization[9,19], phase masks[8], tailored chromatic dispersion[4] and multimode fibers[7,20]. However, previously it has been infeasible to fabricate a needle-beam fiber-optic device without astigmatism for *in vivo* imaging of a blood vessel[8,21-23]. One restriction has been the limited fabrication techniques available for creating complex optical elements, which have made it infeasible to achieve the desired needle-beam shape, while correcting for astigmatism introduced by the catheter sheaths.

Two-photon 3D printing is a high-resolution additive manufacturing process that can fabricate freeform optical structures at the micrometer scale. Preliminary work has demonstrated the potential to correct for astigmatism[3], and fabricate more complex, multi-part optical designs[24].

In this study, we describe the development of a high-resolution, 3D-printed fiber-optic endoscopic probe that utilizes a side-facing axicon lens to acquire 3D images inside a blood vessel. This lens design also corrects for the astigmatism of the protective intracoronary catheter. We characterize the probe, demonstrating a clear superiority in imaging resolution over a standard gradient-index (GRIN) fiber design, and demonstrate its use in both *ex vivo* imaging of atherosclerotic plaques in a human artery, and longitudinal *in vivo* imaging in a porcine model of coronary artery disease.

**Results**

*Freeform needle-beam endoscopic probe design*

Figure 1a depicts the general concept of the needle-beam fiber-optic probe for OCT. The probe was designed using commercial ray-tracing software, Zemax OpticStudio (Ansys, Inc., USA). The goal of the probe design is to enable high-resolution OCT imaging over an extended DOF, for vessels with a diameter of 1–2 mm, while maintaining a small footprint for safe intracoronary imaging. The fiber-optic probe is encapsulated in two tubes: an inner tube and an intracoronary catheter sheath. The inner tube protects the 3D-printed lens as it is inserted into the intracoronary catheter sheath. The intracoronary catheter sheath is then used to enable the placement of the fiber-optic probe within the coronary artery under fluoroscopic and angiographic guidance[25].

Key aspects of the resulting 3D-printed lens design include the mechanism used to redirect the light to achieve side-viewing towards the vessel walls; compensation for astigmatism inherent in this imaging setup; and achieving high lateral resolution over an extended imaging range.

Redirecting the light beam towards the vessel walls is achieved by integrating a reflective surface into the optical design. There are multiple potential approaches to enable this. Recently, a dual-axis catadioptric system was proposed where the mirror to deflect the beam sideways was manufactured by vapor-deposition of a gold-layer[26]. However, this requires additional manufacturing steps, increasing fabrication time and complexity. Our design avoids the need to include a metallic layer, instead we use total internal reflection (TIR) at the air-photoresist interface by ensuring that the incident angle of the light upon the surface is greater than the critical angle of 41°. The reflected light beam propagates at an angle of 80° to the optical axis of the final lens. The output surface of the lens is an axicon with aspheric terms that compensate for irregularities in intensity distribution along the propagation axis. The size of the 3D-printed lens is chosen so that the axicon tip sits extremely close to the inner surface of inner tube (inner diameter ID: 370 µm), minimizing the path length in air to the vessel walls. This 3D-printed axicon forms a needle-beam which then propagates through both catheters at an angle of 10° in the meridional plane.

Secondly, we created a biconical surface which allows it to compensate for the strong astigmatism from two catheters. By compensating for the astigmatism through modifying the lens rather than the shape of the TIR surface, we simplify the reflective surface. This provides a more uniform reflection over a wider range of angles and allows easier optimization of the TIR surface[27].

Finally, after penetrating both catheter sheaths the beam has been optimized assuming that it propagates through 1 mm of blood-like medium before reaching the vessel wall. Figure 1**b** and **c** illustrate the design by microscopic images. The needle-beam starts at the apex of the axicon lens. However, the propagation and shift can be adjusted at the beginning of the needle-beam by tuning the aspheric terms and by controlling the sharpness of the axicon tip.

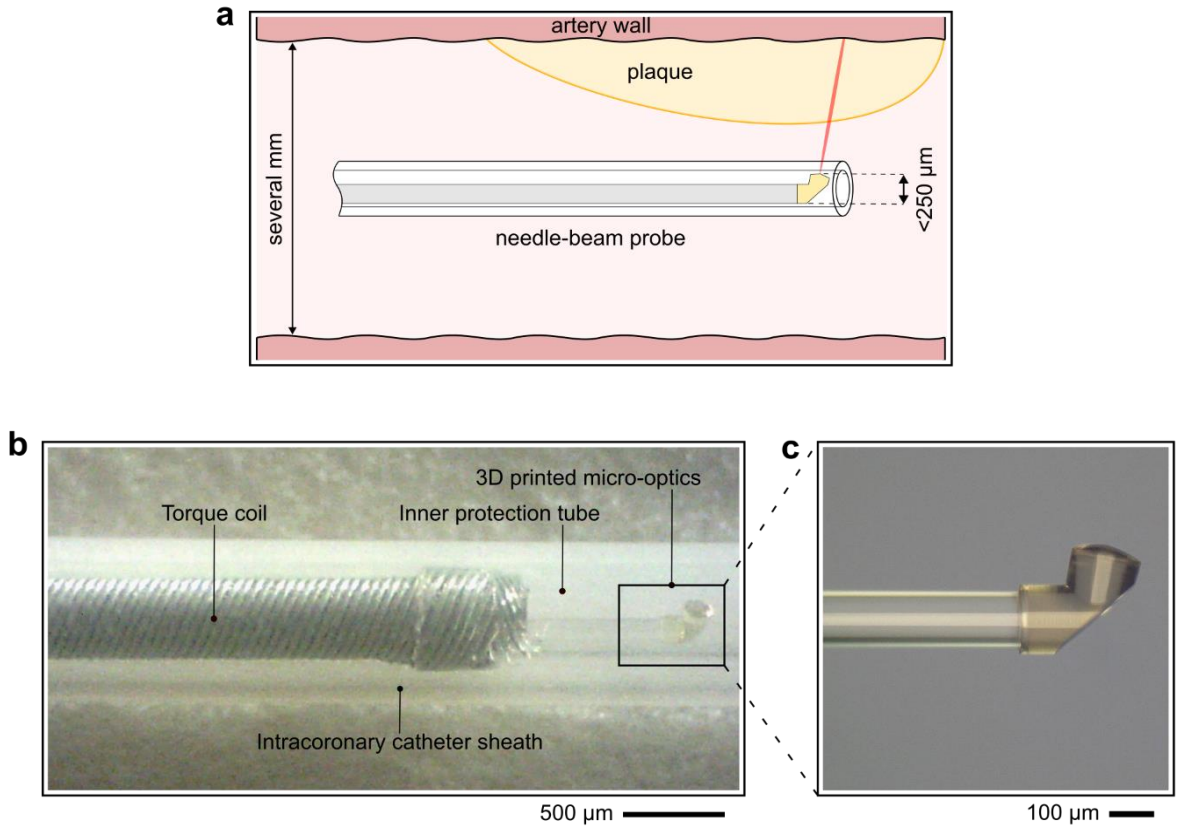

**Figure 1** Freeform needle-beam endoscopic probe design. **a** Sketch of the optical design of the 3D-printed needle-beam probe. Here, we omit mechanical parts which are irrelevant from the optical standpoint. We use a total internal reflectance surface under 80° angle to the optical axis of the fiber together with an axicon to form a needle-beam, which propagates through both catheters and blood/other medium to scan the artery walls. **b** Optical microscope image of the fully assembled needle-beam probe at its distal end. **c** Zoom-in to the 3D-printed axicon micro-optics.

*Characterization of the fabricated needle-beam endoscopic probe*

To evaluate the resolutions ($x$ and $y$ axes) and DOF of the 3D-printed needle-beam probe, we characterized its spatial resolutions as a function of depth in both $x$ and $y$ axes using an OCT resolution phantom (Fig. 2a, APL-OP01, Arden Photonics, UK). Figs. 2b–e display the OCT images acquired by moving the resolution phantom along $x$ and $y$ axes, where $x$ is defined along the axis of the optical fiber, and $y$ is perpendicular to this and to the direction of the light beam. As demonstrated in Figs. 2b and 2d, a nearly astigmatism-free beam was achieved with the 3D-printed needle-beam probe inside the inner tube and a commercial intracoronary catheter sheath. To compare this with a more traditional optical design, we then placed a GRIN fiber probe[28] with both tubes at an equal distance to the resolution phantom. This design of GRIN fiber probe design is commonly used to make OCT endoscopes[29]. The resolution and DOF of the GRIN probe were observed to be inferior to that of the 3D-printed needle beam probe, see Figs. 2c and e. The 3D-printed needle-beam probe can resolve all those bars (11 to 20 μm spacing, i.e., the $m=2$ and $m=3$ vertical lines in Figs. 2b and 2d) for more than 500 μm in depth, while the GRIN fiber probe was unable to resolve any bars between 11 to 20 μm spacing.

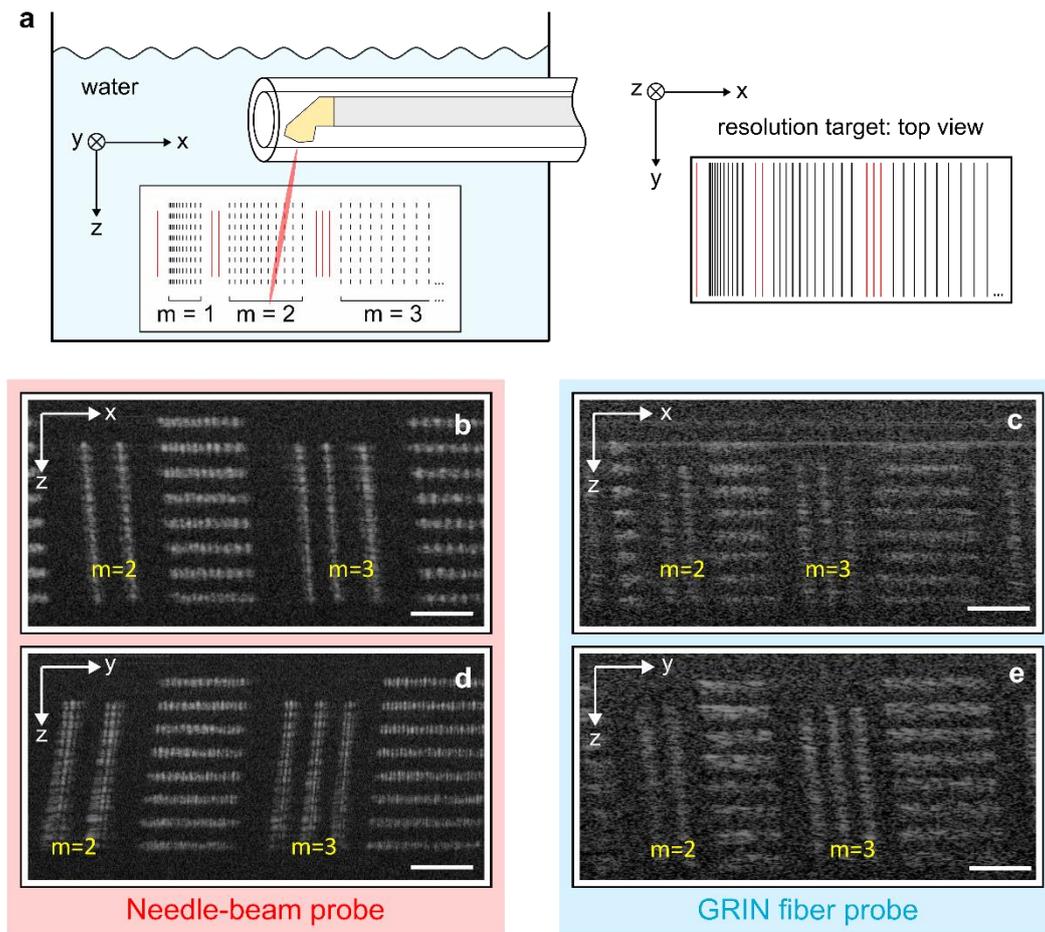

**Figure 2** Resolution characterization. **a** Sketch of the resolution measurement. A resolution target (APL-OP01, Arden Photonics, UK) was pulled back relative to the 3D-printed needle-beam probe along X-axis and later along Y-axis, resulting in two OCT scans – XZ and YZ respectively. The lateral resolution pattern on the APL-OP01 target contains 8 layers. The separation between each subsequent layer is 75 µm (physical distance), so the bottom layer is 525 µm from the top layer (physical distance). The purpose of the lateral resolution pattern is to measure the line spacing. Each line (n) is separated from the next line (n+1) laterally by a distance of 11(m-1)+n where m is the group number. **b** OCT image of the lateral resolution target obtained by the 3D-printed needle-beam probe along the X axis in Fig. 1b; **c** OCT image of the lateral resolution target obtained by a conventional GRIN fiber probe along the X axis; **d** OCT image of the lateral resolution target obtained by the 3D-printed needle-beam probe along the Y axis; **ee** OCT image of the lateral resolution target obtained by a conventional GRIN fiber probe along the Y axis. All images are obtained with the endoscopic probe placed inside the same intracoronary catheter sheath and in water. The 3D-printed lens has a significantly extended depth of focus than that of a GRIN fiber probe. Scale bars indicate optical distances, and the refractive index of the resolution validation phantom is 1.45. Scale bar: 0.25 mm.

*Capability to image human arteries with advanced plaques*

We then imaged *ex vivo* human arteries to explore the improved imaging capability of the 3D-printed needle-beam probe when compared with a conventional OCT GRIN fiber probe. Fig. 3 presents the images obtained in a carotid plaque from a patient who had undergone carotid endarterectomy after suffering a stroke. Figs. 3a and b are the 3D reconstruction of the plaque obtained using the 3D-printed needle-beam endoscopic probe. To ensure that the scans acquired with the 3D-printed and the GRIN fiber probes were co-located for comparison, we held the carotid plaque inside a tube to restrain its movement. When imaged with this 3D-printed needle-beam probe, microstructures in the plaques are visible, as illustrated in Fig. 3c, including acicular (cholesterol) clefts, a feature of high-risk atherosclerotic plaques. The standard GRIN fiber probe (Fig. 3d) was unable to resolve individual clefts. This correlates with our earlier measures of the OCT resolution phantom, as individual cholesterol clefts as typically are approximately 20 µm in size. The existence of cholesterol clefts identified by the 3D-printed probe (Fig. 3c) was validated by the hematoxylin and eosin (H&E) histology image (Fig. 3e). During histological processing, the tube restricting tissue movement was removed to enable histological sectioning. As a result, the lumen of the carotid plaque appears larger in the histology image (Fig. 3e) than those in Figs. 3c and 3d.

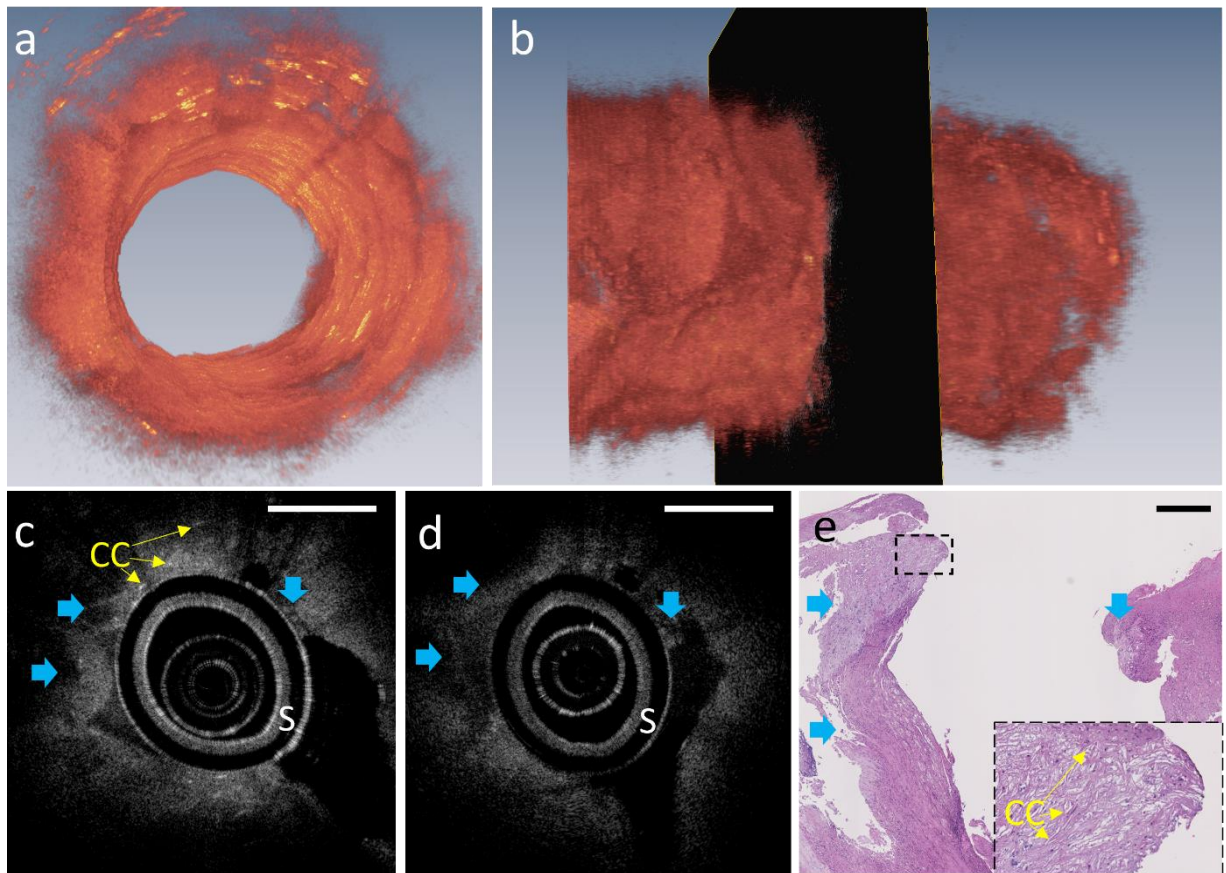

**Figure 3** *Ex vivo* comparison of 3D-printed needle beam probe with conventional OCT probe imaging in a human carotid artery with advanced plaque. **a** and **b** Three-dimensional rendering of the artery created by 210 frames of OCT images obtained with 3D-printed needle-beam endoscopic probe. **c** Representative OCT image obtained at the black box in **b** by the 3D-printed needle-beam endoscopic probe. **d** Representative OCT image obtained at the same location by a GRIN fiber probe. **e** corresponding H&E histology image. Inset: magnified view

of the dashed line region. Blue arrows denote the landmark features used for matching needle-beam, GRIN, and histology images; CC: cholesterol clefts; S: sheaths. Scale bar: 0.5 mm.

*Capability to image inside a coronary artery in vivo*

We performed two *in vivo* pig experiments to assess the capabilities of the 3D-printed needle-beam probe. In the first of these experiments, we conducted intracoronary imaging with both needle-beam and GRIN fiber probes in a live diabetic pig being fed an atherogenic diet for 11 months. This is a previously published pig model of atherosclerosis[30,31]. As demonstrated in each of the sub-images of Figure 4, this coronary artery began to develop an eccentric lesion of neointimal hyperplasia, labelled with an orange arrow at 12 o'clock in the figure. The 3D-printed needle-beam endoscopic probe showed clear differentiation of the neointimal hyperplasia (Fig. 4a). In contrast, the conventional OCT GRIN fiber probe (Fig. 4b) showed poor differentiation of these layered structures and was unable to visualize the neointimal hyperplasia. The lower quality GRIN images are representative of what has previously been reported in the literature[32]. The presence of the neointimal hyperplasia was subsequently validated by H&E histology, as depicted in Fig. 4c.

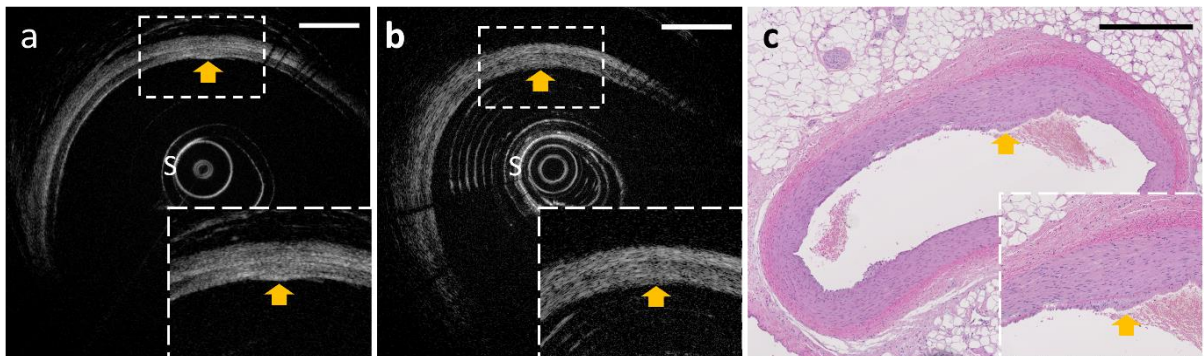

**Figure 4** *In vivo* imaging capability of 3D-printed needle-beam probe in comparison with a conventional GRIN fiber OCT probe. Images of a pig left circumflex coronary artery with early neointimal hyperplasia, highlighted by the orange arrows: **a** image obtained with a 3D-printed needle beam endoscopic probe; **b** image obtained with a conventional GRIN fiber OCT probe; **c** matching H&E histology image confirming the presence of neointimal hyperplasia (orange arrow). S: sheaths. Scale bar: 0.5 mm.

*In vivo intracoronary imaging of a pig across multiple time points*

In a second *in vivo* pig study, we conducted serial imaging of the coronary arteries of a live pig. Images obtained in the left circumflex coronary artery at 3 and 9 months after commencement of the atherogenic diet are shown in Figure 5. These time points allow analysis of two stages of plaque development (early and more advanced)[33,34].

The 3D-printed needle-beam probe was inserted percutaneously through the femoral artery of the pig (Fig. 5a). Under X-ray and angiographic guidance (Fig. 5b), the probe was guided into the coronary arteries, which arise from the aorta and supply the heart with blood. Both the left circumflex and left anterior descending coronary arteries were safely imaged at 50 frames per second by the 3D-printed needle beam probe. The pig successfully recovered from anesthesia after the imaging procedure. A representative image obtained with our 3D-printed needle-beam probe in the left circumflex artery at the 3-month time point is shown in Fig. 5c. It highlights the three-layered arterial structure with intima (i.e., the label 'I' in Fig. 5c), media (i.e., the label 'M', by visualizing the internal elastic lamina 'IEL' between intima and media in Fig. 5c), and adventitial (i.e., the label 'A' in Fig. 5c). The bright spot at 3 o'clock of Fig. 5c (labelled with 'G') is an artefact created by the coronary guide wire, which was used to advance and place the endoscopic probe in the coronary artery. There were no complications after the imaging procedure, and the pig continued the atherogenic diet for another 6 months before repeat surgery and imaging.

At the 9-month time point, the coronary arteries were successfully re-imaged using the 3D-printed needle-beam probe. As displayed in Fig. 5d and 5e, we observed mild progression of the intimal thickening compared to the 3-month timepoint. In the images shown, intimal thickening can be appreciated at both 3-5 o'clock (thicker 'I' layer) and 9-11 o'clock ('P' indicting the position of the atherosclerotic plaque). The presence of plaque was subsequently corroborated with the H&E histology image (Fig. 5e), which was obtained by harvesting the coronaries after the pig was sacrificed following the 9-month time point *in vivo* imaging.

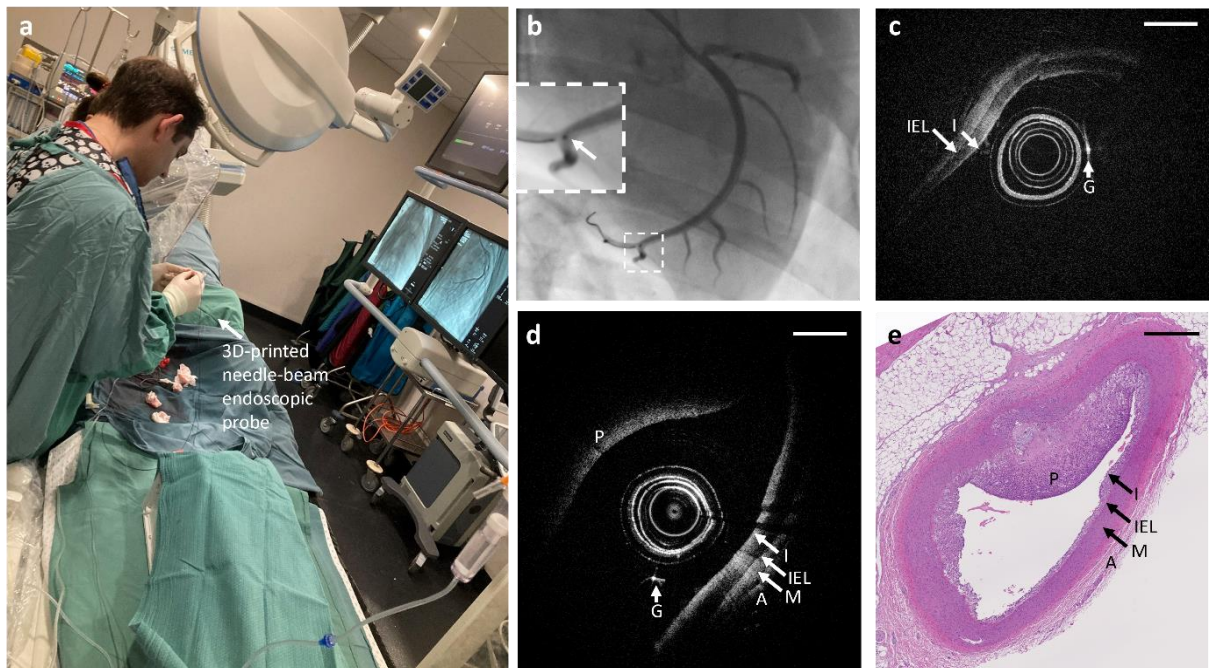

**Figure 5** Serial intracoronary imaging in a live pig with our 3D-printed needle-beam endoscopic probe: **a** Cathlab set up for the intracoronary imaging procedure, where an interventional cardiologist inserted the 3D-printed needle-beam endoscopic probe into the heart and coronary artery of the anaesthetized pig via right femoral artery access; **b** X-ray angiography showing placement of the 3D-printed endoscopic probe in the distal left circumflex artery. Inset: magnified version with an arrow denotes the radiopaque marker at the tip of the 3D-printed needle-beam endoscopic probe; **c** Representative OCT image obtained in the left circumflex coronary artery at the 3-month time point; **d** Representative OCT image obtained in the left circumflex coronary artery at the 9-month time point; **e** matching H&E histological image of the region imaged by **d**. I: intima; IEL: internal elastic lamina; A: adventitia; G: guidewire; P: plaque. Scale bar: 0.5 mm.

## Discussion

To the best of our knowledge, this paper presents the first report of a 3D-printed needle-beam endoscopic probe. This 3D-printed side-viewing freeform lens includes an axicon lens to create a needle beam and a biconic surface to pre-compensate non-chromatic aberration in the system. It demonstrated its ability to obtain 10µm-resolution OCT imaging safely over a large depth range in the torturous arterial systems and coronary arteries in living animals at multiple time points, addressing one of the major limiting features of clinical OCT. The micro-optic design can be easily tailored for each probe, catheter sheath, and application to achieve optimized beam quality and image performance.

Realizing compact fiber-based endoscopes with large DOF and high resolution has previously been technically challenging. Two-photon 3D-printing enables implementing compact but complex optics with various functionalities directly on a fiber tip. Hence, such fibers with 3D-printed micro-optics can be used in OCT-endoscopy applications and may help overcome the current limitations of small optics. The choice of scanning beam shape is a crucial question when designing such micro-optics. As we aim for a large DOF while preserving high resolution, i.e., narrow width of the beam along the axis, needle beams created by axicon-like lenses may provide a solution. Such beams are well-recognized in the biomedical optics field to allow many different approaches for imaging. The resulting needle-like focal shape, extended along its propagation axis, is better suited for achieving the desired characteristics of the beam while preserving rather than compacting the size of the 3D-printed lens.

In our case, we use a 3D-printed axicon lens to create a needle beam over the region of interest along the optical axis, with a typical artery diameter of 2–3 mm. However, the diameter of an artery varies significantly from its proximal to distal end (from approximately 20 mm to few hundred of micrometers). Accordingly, the images obtained by our 3D-printed axicon lens were well-suited for lumen with a small diameter (e.g., Figure 3) but did demonstrate the lens' limitation when the diameter of the artery lumen was larger and the plaque further from the axicon lens (e.g., Figure 5 d). Similarly to the Gaussian or ray optics, the resolution and DOF of the needle beam is dependent on the geometric parameters of the lens, including its' diameter. Therefore, keeping the lens diameter of less than 300 µm while achieving high resolution over a DOF of over 1 mm is still a challenging task.

The 3D micro-printing technique, as demonstrated here, is not only limited to create needle-beam in OCT via an axicon lens, but also other lens/mask/filter[7-10,17-20] for beam shaping/tailoring in various imaging modalities. This technique holds great hope to make beam tailoring in OCT endoscopic probes more precise.

## Methods

### 3D-printed needle beam imaging catheter

Before printing, we prepared the fiber by splicing a no-core fiber (FG125LA, Thorlabs, USA) to a single-mode fiber (SMF28. Thorlabs, USA) using fiber processing machine (Vytran GPX3800, Vytran, UK). With the same machine, we cleaved the no-core fiber to the length of 600 µm. After cleaving, we silanized the fiber probes for at least 2 hours using a solution of 30 ml of ethanol and 150 µl of 3-(trimethoxysilyl)propyl methacrylate to increase the adhesion between the polymer and the fiber facet.

The axicon lens with the TIR surface was printed after the silanization directly onto the fiber facet using Nanoscribe Photonic Professional GT (Nanoscribe GmbH, Germany). We used the commercially available photopolymer IP-S (Nanoscribe GmbH, Germany) to print the structure as this provides the smoothest surface which is crucial to achieve the desired total internal reflection with a reasonable budget for the printing imperfections. We used a slicing size of 0.1 µm and a hatching size of 0.15 µm with a laser power of 60.5% from 50 mW and a scanning speed of 50 mm/s. The needle-beam probe was printed within 3.5 hours. For the development we used a commercially available developer mr-Dev600 (micro resist technology GmbH, Germany) for 15 min and isopropanol for 3 min to rinse the printed structure. Afterwards, we proceeded with 5 min UV-curing with a deep-UV lamp in order to ensure that the structure was fully polymerized, and the refractive index was homogeneous.

The 3D-printed needle-beam fiber probe was placed inside a torque coil (ID of 0.25 mm and OD of 0.5 mm, Asahi Intecc Co.). The distal end of the fiber probe was protected by the inner tube (Zeus, Inc., USA). The protected fiber probe was then inserted into a commercially available intracoronary catheter sheath (Dragonfly, Abbott, USA) for *in vivo* pig experiments. Before each *in vivo* experiment, we sterilized the imaging catheters to be used, in which the 3D-printed needle-beam probes sit, by using ethylene oxide (ETO), a typical way to sterilize medical packaging and device lumens. As long aeration time is needed to remove ETO residue, imaging catheters were given to the animal tech team at least 36 hours before the imaging procedure to ensure sufficient time for sterilization and aeration.

**Back-end imaging system**

The 3D-printed needle-beam imaging catheter was connected to a OCT imaging system. This system used the light source and detector of a commercially available OCT system (with a -5dB bandwidth of the light source ranging from 1185nm to 1415nm and the central wavelength at 1300 nm, Telesto II, Thorlabs, Germany). Between the light source and detector, we connected the intravascular imaging catheter, as the sample arm, and a reference arm consisting of a fiber patch cable and a movable mirror on a motorized linear stage (X-LSM, Zaber, Canada). We tested the capability of scanning with the 3D-printed needle-beam endoscopic probe with two rotation mechanisms: we rotated using a counter-rotation motor[24] and pulled back with a linear stage for the *ex vivo* human study (Fig.3) enabling acquisition of a 3D data volume; For the *in vivo* intracoronary pig study (Figs. 4-5), we utilized a high-speed rotary joint (Princetel, USA) and a linear stage controlled by a MATLAB program (Nine point Medical, USA).

**Ex vivo imaging of human arteries**

The human aspect of this study was approved by Central Adelaide Local Health Network Human Research Ethics Committee (HREC15210). Arteries from patients undergoing standard-of-care clinically indicated carotid or femoral endarterectomy were dissected using standard surgical technique. Instead of discarding the arteries as in the standard protocol, these specimens were scanned by our 3D-printed needle-beam imaging catheter. After imaging, the sample were decalcified by 10% ethylenediaminetetraacetic acid (EDTA) for a week, embedded in paraffin and sectioned for further histology analysis.

**In vivo intracoronary imaging of a pig**

The preclinical part of this study was approved by South Australian Health and Medical Research Institution animal ethics committee (SAM-21-011) and conducted according to the

Australian code for the care and use of animals for scientific purposes (8th ed, 2013; updated 2021). Male Yorkshire swine (35–36 kg) were treated with streptozotocin (50 mg/kg on days 1 and 2, then 100 mg/kg on day 3) to induce diabetes, defined as a consistent blood glucose level >8.3 mmol/L. Starting at the time of diabetes induction, the animals were fed an atherogenic diet. Serum glucose was monitored serially, and insulin were administered to keep glucose levels <30 mmol/L and prevent ketoacidosis. Animal weights and blood glucose levels were monitored monthly. After feeding the animals for more than 3 months, the first intracoronary imaging procedures were conducted. 24-hours prior to imaging, the swine was given a fentanyl patch as pain killer. Prior to the procedure, meloxicam was given via intramuscular (IM) injection at the time of imaging to ensure full pain relief.

On the day of imaging, the animals were sedated with a mixture of Xylazine and Ketamine (IM injection) after which the animal was intubated and ventilated with oxygen (100% v/v). Anaesthesia was maintained by isoflurane inhalation (4-5% for induction, 2-3% for maintenance) with constant oxygen ventilation (2-3 L/hr). Access to the porcine coronary system was obtained percutaneously via a femoral artery. A standard 6F coronary guiding catheter (AL2) was introduced via a 6F sheath into one of the three main coronary arteries under X-ray and angiographic guidance. The 3D-printed needle-beam endoscopic probe was advanced into the coronary artery via the guiding catheter. Heparin was administered during procedure to prevent thrombus formation during the procedure and has a half-life of approximately 1 hour. Following the imaging procedure, all catheters were removed while the animals are under anesthesia, and the sheath removed, and hemostasis obtained by manual pressure. Antibiotics were used as required under veterinary guidance.

The animals were closely monitored and continuously fed with an atherogenic diet for another 6 months after the first imaging procedure. The terminal imaging procedures were conducted similar to the protocol described above via right femoral artery access. The 3D-printed needle-beam endoscopic probe was advanced into the same coronary arteries as imaged at the 3-month time point by using X-ray angiography guidance. At the conclusion of the study, animals were euthanized, and their coronary arteries were processed for histology.

**Acknowledgements**

We acknowledge Dr. Chris Christou, Ms. Loren Matthews, Ms. Georgia Williams, Ms. Lisa McKenny, Mr. Robb Muirhead, Dr Albert Kota, Dr Christina Popovic, Mr. Jim Manavis, Ms. Sofie Kogoj, Ms. Yvonne Ciuk, A/Prof John Finnie, Prof Alois Herkommer for contributing to the management of the diabetic pigs, assistance during imaging procedures, collection of human carotid arteries, histological preparation, analysis of histology images and discussion about 3D-printing designs. The authors wish to thank individuals who donated their tissues for the advancement of education and research. They also acknowledge the support of the Optofab node of the Australian National Fabrication Facility and Adelaide Microscopy.

**Funding**

The authors have been supported by the National Health and Medical Research Council (NHMRC) Development grant (2022337), Ideas Grant (2001646), and Investigator Grant (2008462), the Heart Foundation Future Leader Fellowship (105608 and 106656), the hospital research foundation project grant (2022-CP-IDMH-014-83100), Australia-Germany Joint Research Co-operation Scheme (UA-DAAD), Baden-Wuerttemberg-Stiftung (Opterial), European Research Council (Advanced Grant Complexplas, PoC Grant 3DPrintedOptics), Bundesministerium für Bildung und Forschung (3DprintedOptics, Integrated3Dprint),

Deutsche Forschungsgemeinschaft (DFG, German Research Foundation), European Union (IEC IV-Lab), Carl-Zeiss foundation (Endoprint3D), and University of Stuttgart (Terra Incognit).

**Conflict of interest**

P.J.P. has received research support from Abbott Vascular, consulting fees from Amgen, Esperion, Eli Lilly, Novartis, Novo Nordisk and Sanofi, and speaker honoraria from Amgen, AstraZeneca, Bayer, Boehringer Ingelheim, Merck Schering-Plough, Pfizer, Novartis, Novo Nordisk and Sanofi. R.A.M. is a co-founder and Director of Miniprobes Pty Ltd, a company that develops optical imaging systems. Miniprobes Pty Ltd did not contribute to or participate in this study. The remaining authors declare no competing interests.